\documentclass[10pt,superscriptaddress,prc,aps,twocolumn,showpacs]{revtex4}
\usepackage{amsmath,amssymb}
\usepackage{graphicx}
\usepackage{hyperref} 
\newcommand{\Mainz}[1]
{\affiliation{Institut f\"ur Kernphysik, Johannes Gutenberg-Universit\"at Mainz, D-55099 
Mainz,Germany}}

\newcommand{\Bonn}[1]
{\affiliation{Helmholtz-Institut f\"ur Strahlen- und Kernphysik, Universit\"at Bonn, D-53115 
Bonn, Germany}}

\newcommand{\Kent}[1]
{\affiliation{Kent State University, Kent, Ohio 44242-0001, USA}}

\newcommand{\Glasgow}[1]
{\affiliation{SUPA School of Physics and Astronomy, University of Glasgow, Glasgow G12 8QQ, 
United Kingdom}}

\newcommand{\Pavia}[1]
{\affiliation{INFN Sezione di Pavia, I-27100 Pavia, Italy}}

\newcommand{\GWU}[1]
{\affiliation{The George Washington University, Washington, DC 20052-0001, USA}}

\newcommand{\LPI}[1]
{\affiliation{Lebedev Physical Institute, 119991 Moscow, Russia}}

\newcommand{\Dalhousie}[1]
{\affiliation{Dalhousie University, Halifax, Nova Scotia B3H 4R2, Canada}}

\newcommand{\Halifax}[1]
{\affiliation{Department of Astronomy and Physics, Saint Mary’s University, Halifax, Nova 
Scotia B3H 3C3, Canada}}

\newcommand{\UniPavia}[1]
{\affiliation{Dipartimento di Fisica, Universit\`a di Pavia, I-27100 Pavia, Italy}}

\newcommand{\Basel}[1]
{\affiliation{Departement f\"ur Physik, Universit\"at Basel, CH-4056 Basel, Switzerland}}

\newcommand{\Tomsk}[1]
{\affiliation{Laboratory of Mathematical Physics, Tomsk Polytechnic University, 634034 
Tomsk, Russia}}

\newcommand{\Edinburgh}[1]
{\affiliation{SUPA School of Physics, University of Edinburgh, Edinburgh EH9 3JZ, United 
Kingdom}}

\newcommand{\INR}[1]
{\affiliation{Institute for Nuclear Research, 125047 Moscow, Russia}}

\newcommand{\Sackville}[1]
{\affiliation{Mount Allison University, Sackville, New Brunswick E4L 1E6, Canada}}

\newcommand{\Regina}[1]
{\affiliation{University of Regina, Regina, Saskatchewan S4S 0A2, Canada}}

\newcommand{\Zagreb}[1]
{\affiliation{Rudjer Boskovic Institute, HR-10000 Zagreb, Croatia}}

\newcommand{\Amherst}[1]
{\affiliation{University of Massachusetts, Amherst, Massachusetts 01003, USA}}

\newcommand{\Bochum}[1]
{\affiliation{Institut f\"ur Experimentalphysik, Ruhr-Universit\"at , D-44780 Bochum, 
Germany}}

\newcommand{\UCLA}[1]
{\affiliation{University of California Los Angeles, Los Angeles, California 90095-1547, 
USA}}

\begin{document}

\title{
Study of $\eta$ and $\eta'$ photoproduction at MAMI}
\author{V.~L.~Kashevarov}\Mainz \\ \LPI \\
\author{P.~Ott}\Mainz \\
\author{S.~Prakhov}\Mainz \\ \GWU \\ \UCLA \\
\author{P.~Adlarson}\Mainz \\
\author{F.~Afzal}\Bonn \\
\author{Z.~Ahmed}\Regina \\
\author{C.~S.~Akondi}\Kent \\
\author{J.~R.~M.~Annand}\Glasgow  \\
\author{H.~J.~Arends}\Mainz \\
\author{R.~Beck}\Bonn \\
\author{A.~Braghieri}\Pavia \\
\author{W.~J.~Briscoe}\GWU \\
\author{F.~Cividini}\Mainz \\
\author{R.~Codling}\Glasgow \\
\author{C.~Collicott}\Dalhousie \\ \Halifax \\
\author{S.~Costanza}\Pavia \\ \UniPavia \\
\author{A.~Denig}\Mainz \\
\author{E.~J.~Downie}\Mainz \\ \GWU \\
\author{M.~Dieterle}\Basel \\
\author{M.~I.~Ferretti Bondy}\Mainz \\
\author{L.~V.~Fil'kov}\LPI \\
\author{A.~Fix}\Tomsk \\
\author{S.~Gardner}\Glasgow \\
\author{S.~Garni}\Basel \\
\author{D.~I.~Glazier}\Glasgow \\ \Edinburgh \\
\author{D.~Glowa}\Edinburgh \\ 
\author{W.~Gradl}\Mainz \\
\author{G.~Gurevich}\INR \\
\author{D.~J.~Hamilton}\Glasgow \\
\author{D.~Hornidge}\Sackville \\
\author{D.~Howdle}\Glasgow \\
\author{G.~M.~Huber}\Regina \\
\author{A.~K\"aser}\Basel\\  
\author{S.~Kay}\Edinburgh \\  
\author{I.~Keshelashvili}\Basel\\
\author{R.~Kondratiev}\INR \\
\author{M.~Korolija}\Zagreb \\
\author{B.~Krusche}\Basel \\ 
\author{J.~Linturi}\Mainz \\
\author{V.~Lisin}\LPI \\
\author{K.~Livingston}\Glasgow \\
\author{I.~J.~D.~MacGregor}\Glasgow \\
\author{R.~MacRae}\Glasgow \\
\author{J.~Mancell}\Glasgow \\
\author{D.~M.~Manley}\Kent \\
\author{P.~P.~Martel}\Mainz \\ \Sackville \\
\author{J.~C.~McGeorge}\Glasgow \\
\author{E.~McNicol}\Glasgow \\
\author{D.~G.~Middleton}\Mainz \\ \Sackville \\
\author{R.~Miskimen}\Amherst \\
\author{E.~Mornacchi}\Mainz \\
\author{C.~Mullen}\Glasgow \\
\author{A.~Mushkarenkov}\Pavia \\ \Amherst \\
\author{A.~Neiser}\Mainz \\
\author{M.~Oberle}\Basel \\
\author{M.~Ostrick}\thanks{ostrick@kph.uni-mainz.de}\Mainz \\  
\author{P.~B.~Otte}\Mainz \\ 
\author{B.~Oussena}\Mainz \\ \GWU \\
\author{D.~Paudyal}\Regina \\ 
\author{P.~Pedroni}\Pavia \\
\author{V.~V.~Polyanski}\LPI \\
\author{A.~Rajabi}\Amherst \\ 
\author{G.~Reicherz}\Bochum \\
\author{J.~Robinson}\Glasgow \\
\author{G.~Rosner}\Glasgow \\
\author{T.~Rostomyan}\Basel \\
\author{A.~Sarty}\Halifax \\
\author{D.~M.~Schott}\GWU \\
\author{S.~Schumann}\Mainz \\
\author{C.~Sfienti}\Mainz \\
\author{V.~Sokhoyan}\Mainz \\ \GWU \\
\author{K.~Spieker}\Bonn \\
\author{O.~Steffen}\Mainz \\  
\author{B.~Strandberg}\Glasgow \\
\author{I.~I.~Strakovsky}\GWU \\
\author{Th.~Strub}\Basel \\
\author{I.~Supek}\Zagreb \\
\author{M.~F.~Taragin}\GWU \\
\author{A.~Thiel}\Bonn \\  
\author{M.~Thiel}\Mainz \\ 
\author{L.~Tiator}\Mainz \\ 
\author{A.~Thomas}\Mainz \\
\author{M.~Unverzagt}\Mainz \\
\author{S.~Wagner}\Mainz \\   
\author{D.~P.~Watts}\Edinburgh \\
\author{D.~Werthm\"uller}\Glasgow \\ \Basel \\
\author{J.~Wettig}\Mainz \\
\author{L.~Witthauer}\Basel \\
\author{M.~Wolfes}\Mainz \\   
\author{R.~L.~Workman}\GWU \\ 
\author{L.~Zana}\Edinburgh \\

\collaboration{A2 Collaboration at MAMI}

\date{\today}

\begin{abstract}
 The reactions $\gamma p\to \eta p$ and $\gamma p\to \eta' p$
 have been measured from their thresholds up to the center-of-mass energy $W=1.96$~GeV
 with the tagged-photon facilities at the Mainz Microtron, MAMI.
 Differential cross sections were obtained with unprecedented accuracy,
 providing fine energy binning and full production-angle coverage. 
 A strong cusp is observed in 
 the total cross section and excitation functions for $\eta$ photoproduction
 at the energies in vicinity of the $\eta'$ threshold, $W=1896$~MeV ($E_\gamma=1447$~MeV).
 This behavior is explained in a revised $\eta$MAID isobar model
 by a significant branching of the $N(1895)1/2^-$ nucleon resonance to both, 
 $\eta p$ and $\eta' p$, confirming the existence and 
 constraining the properties of this poorly known state.  
 \end{abstract}


\maketitle

 
The photo-induced production of $\eta$ and $\eta'$ mesons is a selective probe 
to study excitations of the nucleon. The $\eta$ and the $\eta'$ represent the isoscalar members of the fundamental
pseudoscalar-meson nonet and, in contrast to the isovector $\pi$, excitations with isospin $I = 3/2$ ($\Delta$ resonances)
do not decay into $\eta N$ and $\eta' N$ final states. 
Several single and double-spin observables of the $\gamma p \to \eta p$ reaction have recently been measured 
\cite{CLAS_2016, A2MAMI_2014, McNicoll2010, CLAS_2009, CBELSA_2009, GRAAL_2007}. 
A review of the experimental and phenomenological progress can be found in Ref.~\cite{Krusche2014}. 
All model calculations 
\cite{BnGa11, SAID, Giessen2012, Kamano2013, BnJue, etaMAID_2003} agree in the dominance of the $E_{0+} (J^P = 1/2^-)$ multipole amplitude, 
which is populated by the well established $N(1535)1/2^-$ and $N(1650)1/2^-$ resonances. 
The existence of a third $1/2^-$ nucleon 
resonance, however, is still under discussion. The $N(1895)1/2^-$ is presently listed by the PDG with only 
two stars \cite{PDG}. 
 The experimental data for $\eta'$ production is much more scarce. 
 The most recent measurements 
 by CLAS~\cite{CLAS_2006, CLAS_2009} and CBELSA/TAPS~\cite{CBELSA_2009} decreased
 uncertainties in the $\gamma p \to \eta' p$ differential cross sections,
 leaving, however, the near-threshold region still unexplored.
 Recently, this threshold region attracted additional
 attention, after the first results for the beam asymmetry $\Sigma$
 were presented by GRAAL~\cite{GRAAL_2015} which, although limited in statistics, could not be reproduced by any of 
 the existing models describing $\eta'$ photoproduction \cite{Nakayama_2006,Nakayama_2013,etaMAIDr_2003,Zhong_2011,Tryasuchev_2013}.
 The threshold for the $ \gamma p \to \eta' p$ reaction at $W=1896$~MeV is located in a mass 
region that plays a key role for our understanding of the nucleon spectrum.
Presently, there are no well established (four stars) states
 between $W = 1800-2100$~MeV. However, there are many state candidates
 and an even larger number of states predicted by quark-models~\cite{CapRob, Bonn} or 
lattice QCD \cite{Lattice}. 

 This work contributes to the study of $\eta$ and $\eta'$ photoproduction
 by presenting new, high-statistics measurements of the $\gamma p\to\eta p$
 and $\gamma p\to\eta' p$ differential cross sections 
 from reaction thresholds up to $E_\gamma=1577$~MeV
 ($W=1960$~MeV). The data were obtained with a fine
 binning in $E_\gamma$ and cover the full range of the production angles.

The experiments were conducted using the Crystal Ball (CB)~\cite{CB}
as a central calorimeter and TAPS~\cite{TAPS}
as a forward calorimeter. These detectors were
installed at the energy-tagged bremsstrahlung-photon beam
produced from the electron beam of the Mainz Microtron (MAMI)~\cite{MAMIC}.
The beam photons were incident on a liquid hydrogen target located
in the center of the CB. The energies of bremsstrahlung photons, $E_\gamma$,
produced by the electrons in a $10\,\mu$m copper radiator, were analyzed
by detecting postbremsstrahlung electrons in tagging spectrometers (taggers).
The Glasgow-Mainz tagger\,\cite{TAGGER} was used in the major part of the experiments. 
In order to tag the high-energy part of the bremsstrahlung spectrum, a dedicated  end-point tagging 
spectrometer (EPT)~\cite{pi0_a2_2015} was used, especially
designed for $\eta'$ measurements.

In this letter, we present the analysis of three independent data sets from
different periods of data taking.
The first data set (Run-I) was taken in 2007 with the 1508-MeV electron beam
and the bremsstrahlung photons analyzed by the Glasgow-Mainz tagger up to
an energy of 1402~MeV. All details on the experimental resolution
 of the detectors and other conditions during these measurements
 are given in Refs.~\cite{etaslope2009,McNicoll2010} and references therein.
 In Ref.~\cite{McNicoll2010}, the total and differential cross sections for 
 the $\gamma p \to \eta p$ reaction were obtained by identifying the $\eta$ meson
 via its $3 \pi^0$ decay mode. 
 This analysis was repeated with an improved cluster algorithm,
 better separating electromagnetic showers partially overlapping
in the calorimeters. The second important neutral 
 decay mode $\eta \to \gamma \gamma$ was analyzed as well.
  The second data set (Run-II) was taken in 2009
 with the 1557-MeV electron beam and the bremsstrahlung photons analyzed
 up to 1448~MeV. 
 The trigger conditions for this run required more than two clusters
to be detected in the CB, which suppressed severely the detection of
 $\eta \to \gamma \gamma$ decays, and only $\eta\to 3\pi^0$ decays were reconstructed
 in the analysis. More details on the Run-II conditions can be found in
 Ref.~\cite{K0Sigpl2013}.    
  The third data set (Run-III) was taken in 2014 with the 1604-MeV electron beam
 and the bremsstrahlung photons analyzed by the EPT spectrometer
 from 1426~MeV up to 1576~MeV. 
 In this run, the energy of the $\eta'$ production threshold
was covered, and both neutral $\eta$ decay modes  as well as 
the $\eta'\to \gamma\gamma$ and $\eta'\to \pi^0\pi^0\eta\to 6\gamma$ decays  
 were reconstructed. More details on the Run-III conditions can be
 found in Ref.~\cite{pi0_a2_2015}. 


 The selection of event candidates and the reconstruction of the outgoing particles 
 was based on the kinematic-fit technique.
 Details on the kinematic-fit parametrization of the detector
 information and resolutions are given in Ref.~\cite{etaslope2009}.
 The determination of the experimental acceptance for each decay mode of $\eta$ and $\eta'$
 was based on a Monte Carlo (MC) simulation of all processes $\gamma p\to \eta^{(')} p$.
 The generated events were propagated through a {\sc GEANT} simulation of the experimental
 setup. To reproduce resolutions of the experimental data, the {\sc GEANT}
 output was subject to additional smearing, thus allowing both the simulated
 and experimental data to be analyzed in the same way.
\begin{figure}[h]
\begin{center}
\resizebox{0.48\textwidth}{!}{\includegraphics{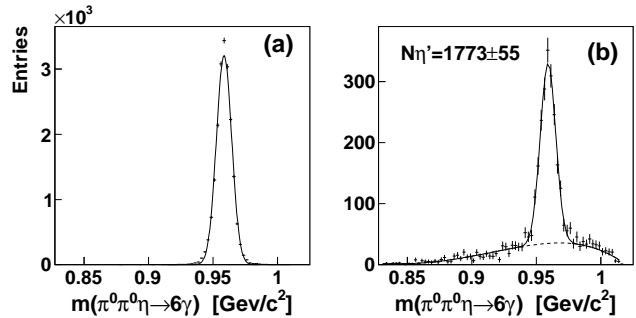}}
\caption{
$m(\pi^0\pi^0\eta\to 6\gamma)$ invariant-mass distributions obtained for $E_\gamma=1558$~MeV 
and $\cos\theta=0.1\pm 0.01$:
 (a)~MC simulation of $\gamma p\to \eta' p\to \pi^0\pi^0\eta p\to 6\gamma p$ with a Gaussian fit;
 (b)~experimental spectrum fitted with the sum of a Gaussian and a polynomial of order four.
}
 \label{fig1}
\end{center}
\end{figure}

 A possible background was investigated via Monte Carlo simulation of competing reactions.
 For both the decay modes of $\eta'$ no background sources were found that 
 could produce a peak in the $m(\gamma\gamma)$ and $m(\pi^0\pi^0\eta\to 6\gamma)$ invariant-mass distributions at the position of the $\eta'$ mass.
 However, the selection of event candidates with the kinematic fit was not
 sufficient to eliminate all background in vicinity of the $\eta'$.
 Thus, the number of $\eta'$ decays observed in every energy-angle bin
 was obtained by fitting experimental $m(\gamma\gamma)$
 and $m(\pi^0\pi^0\eta\to 6\gamma)$ spectra with a function,
 describing the $\eta'$ peak above a smooth background.
This procedure is illustrated for one energy-angle bin
 in Fig.~\ref{fig1}, showing a typical $\eta'\to \pi^0\pi^0\eta\to 6\gamma$ 
 invariant-mass distribution and the background shape.
 In total, all selected events were divided into 10 $\cos\theta$ bins,
 where $\theta$ is the meson production angle in the c.m.\ frame.
 The covered energy range, $E_\gamma=$1447--1577~MeV, was divided
 into 12 intervals, with the first four 6.5-MeV wide and next
 eight 13-MeV wide.
 
 For the $\gamma p\to \eta p$ differential cross sections,
 all selected events were divided into 24 $\cos\theta$ bins.
 For energies below $E_\gamma=1.25$~GeV, the present analysis of
 the process $\gamma p\to \eta p\to 3\pi^0 p\to 6\gamma p$ was very
 similar to the method described in detail in Ref.~\cite{McNicoll2010}.
 At higher energies, as in the case of $\eta'$, the background under the 
 $\eta$ decays 
 could not be fully eliminated, and the same fitting
procedure, as described above for $\eta'$, was applied.
  
  The $\gamma p\to \eta p$ and $\gamma p\to \eta' p$ differential cross sections
 were obtained by taking into account the values for the corresponding
 $\eta$ and $\eta'$ branching ratios~\cite{PDG},
 the number of protons in the hydrogen target,
 and the photon-beam flux from the tagging facilities, corrected by the
 fraction rejected by the collimator.
 For the $\eta$ cross sections, the overall systematic uncertainty due to
 the calculation of the detection efficiency and the photon-beam flux
 was estimated similar to our previous analyses  \cite{McNicoll2010, A2MAMI_2014} 
 as 4\% for the data taken in Run-I and Run-II,
 and as 5\% for the data taken in Run-III. Similar systematic uncertainty
 for the $\eta'$ cross sections from Run-III is also 5\%.
\begin{figure}
\begin{center}
\resizebox{0.46\textwidth}{!}{\includegraphics{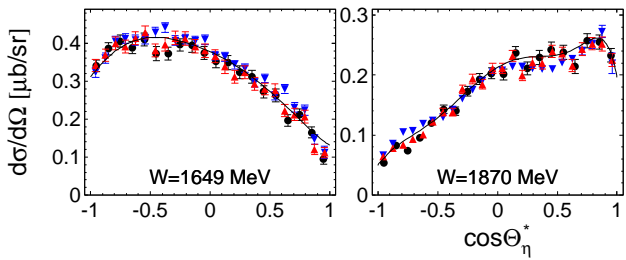}}
\caption{(Color online)
 Comparison of the $\gamma p\to \eta p$ differential
 cross sections from Run-I (red triangles for the $\eta \to \gamma \gamma$ and blue 
 triangles for the $\eta \to 3 \pi^0$ decay mode) with
 the previous analysis of Ref.\,\protect\cite{McNicoll2010} 
 (black circles, $\eta \to 3\pi^0$ decay)
for selected energy bins.
 The error bars of all data points represent statistical uncertainties only.
 The line shows the new $\eta$MAID2016 solution discussed in the text.
}
\label{fig2}
\end{center}
\end{figure}
 In Fig.~\ref{fig2}, the results from Run-I for both decay modes are
 compared to the previous analysis of Ref.~\cite{McNicoll2010}.
\begin{figure}
\begin{center}
\resizebox{0.46\textwidth}{!}{\includegraphics{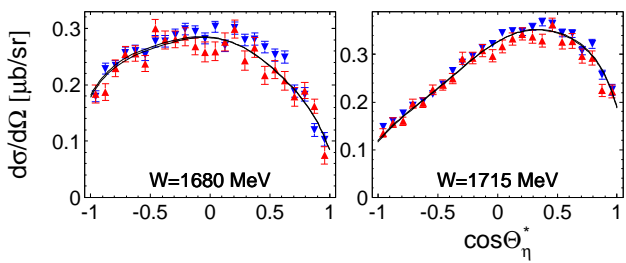}}
\caption{(Color online)
 Comparison of the $\gamma p\to \eta p$ differential
 cross sections for Run-I (blue) and 
  Run-II (red) in selected energy bins where the largest discrepancies are
  observed. 
  The line shows the new $\eta$MAID2016 solution.  
}
\label{fig2b}
\end{center}
\end{figure}
A comparison of the differential cross sections from Run-I and Run-II for two 
selected energy bins, where the largest discrepancies are observed,
is illustrated in Fig.~\ref{fig2b}. 
Finally, Fig.~\ref{fig3} checks Run-II against Run-III, which used a 
different tagging spectrometer.   
\begin{figure}
\begin{center}
\resizebox{0.46\textwidth}{!}{\includegraphics{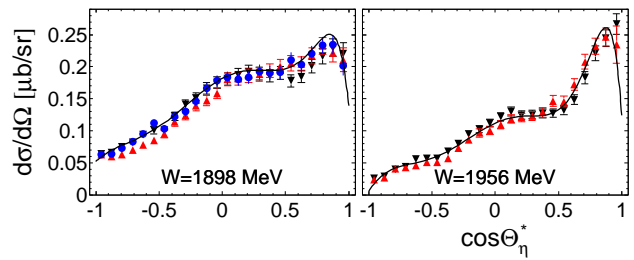}}
\caption{(Color online)
$\gamma p\to \eta p$ differential cross sections for minimal and maximal energies
measured during Run-III: the black triangles ($\eta \to 2\gamma$), and red triangles ($\eta \to 3\pi^0$) were obtained in Run-III. The blue circles show results from Run-II ($\eta \to 3\pi^0$).
The line shows the new $\eta$MAID2016 solution. 
}
\label{fig3}
\end{center}
\end{figure}
In general, the different data sets are in agreement within the given 
uncertainties. 
 To reflect small discrepancies, which can be observed in particular  
regions with larger background,  an additional 3\% systematic uncertainty 
reflecting uncertainties in the angular dependence of the reconstruction efficiency
 was added in quadrature to all statistical uncertainties
 in the $\eta \to \gamma\gamma$ and  $\eta\to 3\pi^0$ results of Run-I
 and Run-II above $E_\gamma=1.25$~GeV, and 5\% for Run III.
 These uncertainties were then used to combine the $\eta \to \gamma\gamma$
 and $\eta\to 3\pi^0$ results together. 
 Similar systematic uncertainties were estimated as 5\% for $\eta' \to \gamma\gamma$
 and 6\% for $\eta'\to \pi^0\pi^0\eta\to 6\gamma$.
The agreement of our $\gamma p\to \eta p$ differential cross section measurements with previous data 
was already demonstrated in \cite{McNicoll2010}.
At high energies, the results from CLAS \cite{CLAS_2009} are in a better agreement
with our present data than those from CBELSA/TAPS \cite{CBELSA_2009}. 
\begin{figure}
\begin{center}
\resizebox{0.46\textwidth}{!}{\includegraphics{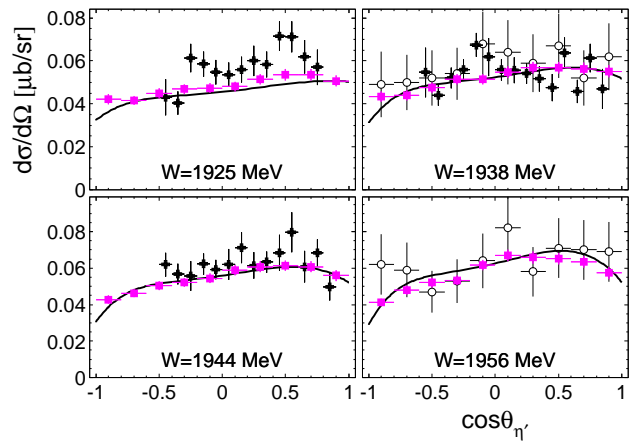}}
\caption{(Color online)
Present results for the
 $\gamma p\to \eta' p$ differential cross sections (magenta squares) compared 
 to previous measurements
 (black crosses for CLAS~\protect\cite{CLAS_2009} and black open circles
 for CBELSA/TAPS~\protect\cite{CBELSA_2009}) and to the new $\eta$MAID2016 solution
 (black solid line). 
 }
\label{fig4}
\end{center}
\end{figure}
 The new results for the $\gamma p\to \eta' p$ differential cross sections
 are illustrated in Fig.~\ref{fig4} for four energy bins which
 overlap with the data from CLAS~\cite{CLAS_2009} and
 CBELSA/TAPS~\cite{CBELSA_2009}. Our results are in agreement with the previous 
 data within the error bars, but have a much superior statistical accuracy.

\begin{figure}
\begin{center}
\resizebox{0.46\textwidth}{!}{\includegraphics{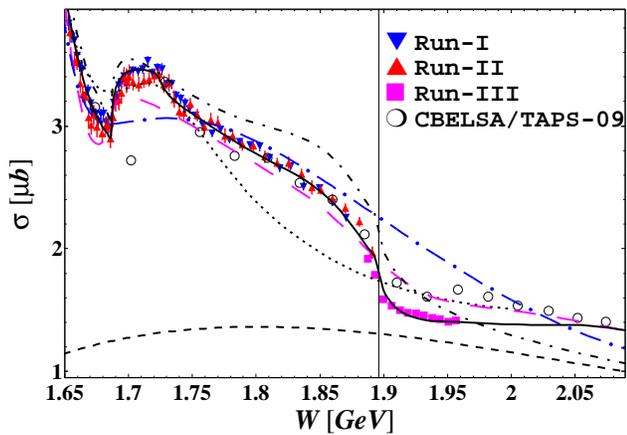}}
\caption{(Color online)
Total cross section obtained for the $\gamma p\to \eta p$ reaction
along with previous measurements by CBELSA/TAPS~\protect\cite{CBELSA_2009}.
The data are compared to the model calculations $\eta$MAID-2003~\protect\cite{etaMAID_2003} 
(black dotted line), SAID-GE09~\protect\cite{SAID} (blue long dashed-dotted line), and
BG2014-2~\protect\cite{BnGa11} (magenta long dashed line). The new  
$\eta$MAID2016 solution is shown as a black solid line. The Regge-background (dashed) 
as well as the sum of background and the contributions from N$1/2^-$-resonances (dashed-dotted)
are shown separately.
 }
\label{fig5eta}
\end{center}
\end{figure}

\begin{figure}
\begin{center}
\resizebox{0.46\textwidth}{!}{\includegraphics{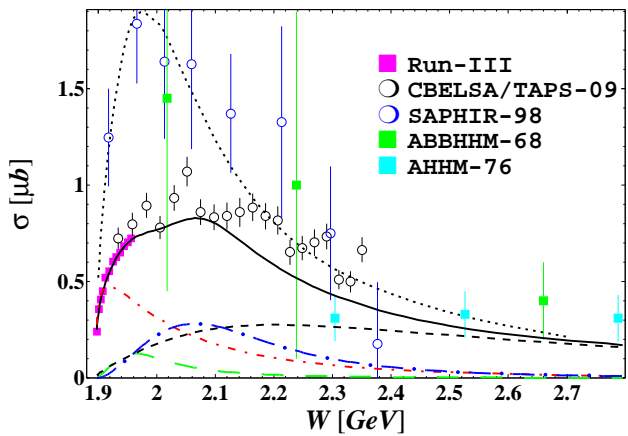}}
\caption{(Color online)
Total cross section obtained for the  $\gamma p\to \eta' p$ 
reaction along with previous measurements by CBELSA/TAPS~\protect\cite{CBELSA_2009}, 
 SAPHIR~\protect\cite{SAPHIR_1998},
ABBHHM~\protect\cite{ABBHHM_1968}, and AHHM~\protect\cite{AHHM_1976}.
The data are compared to the model calculations  
$\eta$MAID-Regge-2003~\protect\cite{etaMAIDr_2003} (black dotted line) and to 
the new  $\eta$MAID2016 solution (black solid line). 
Furthermore,  the background (black dashed) as well as the main resonance
contributions $N(1895)1/2^-$ (red dashed-dotted), $N(1990)1/2^+$ (green long dashed), and $N(2020)3/2^-$ (blue long dashed-dotted)
are shown separately.
 }
\label{fig5etapr}
\end{center}
\end{figure}
The total cross sections 
were obtained by integrating the corresponding differential cross sections.
The results obtained for the $\gamma p \to \eta p$ and  $\gamma p \to \eta' p$ reactions
are shown in  Fig.~\ref{fig5eta} and Fig.~\ref{fig5etapr}, respectively.
The comparison with previous data in the figures clearly demonstrates the high accuracy of our 
new measurements.

\begin{figure*} 
\begin{center}
\resizebox{1.0\textwidth}{!}{\includegraphics{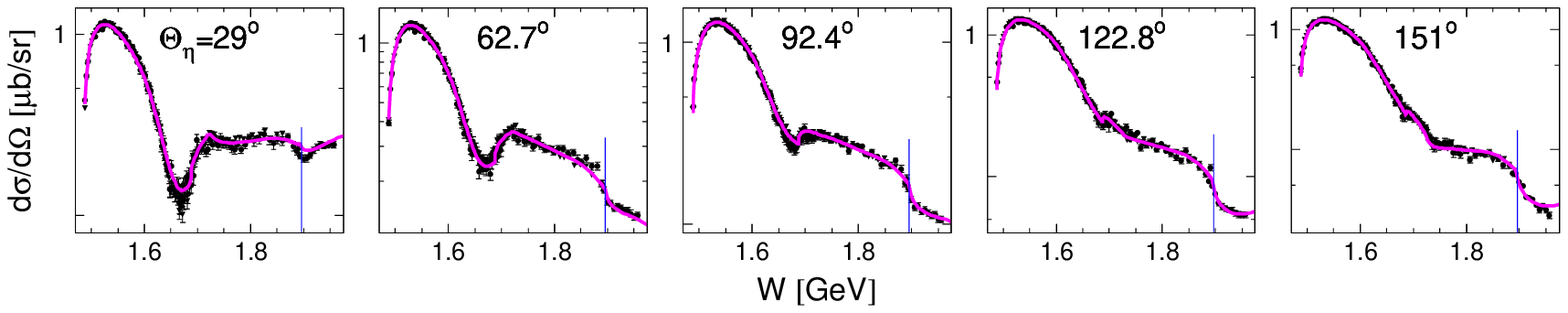}}
\caption{(Color online)
Excitation function of $\eta$ photoproduction for selected angular bins. Black circles are the present 
data, the magenta line is the new $\eta$MAID2016 solution, the vertical line corresponds to the $\eta'$ threshold. 
}
\label{fig6}   
\end{center}
\end{figure*}
Besides the distinct dip at $W=1670$~MeV \cite{McNicoll2010}, our new data for the $\gamma p\to \eta p$ 
reaction show another pronounced feature at higher energies. At the position
of the $\eta'$ threshold at $W=1896$~MeV, marked by the  vertical line in Fig.~\ref{fig5eta},  
a clear cusp is observed.
The sharpness of this cusp is strongly dependent on the polar angle of the $\eta$ meson, as shown in Fig.~\ref{fig6}. 
While the dip at $W = 1670$~MeV is more pronounced at forward angles, the cusp effect is stronger around $90^o$.


One of the first dedicated models for photoproduction of $\eta$ and $\eta'$ mesons
was the Mainz isobar model $\eta$MAID~\cite{etaMAID_2003, etaMAIDr_2003}, which was
fitted to data available in 2003. These fits are shown as dotted lines 
in Figs.~\ref{fig5eta} and ~\ref{fig5etapr},
and there is no surprise that they fail to reproduce the current measurements.  
However, even the more recent analyses by SAID-GE09~\cite{SAID} and BG2014-2~\cite{BnGa11}, 
are still far from agreement with the new precision data.

To interpret the new data we have developed a model based on the ideas of
$\eta$MAID \cite{etaMAID_2003, etaMAIDr_2003}.    
This new $\eta$MAID2016 model includes a non-resonant background, which consists of 
the vector ($\rho$ and $\omega$) and
axial-vector ($b_1$) exchange in the $t$ channel, and $s$-channel resonance excitations. Regge trajectories
for the meson exchange in the $t$ channel were used to provide correct asymptotic behavior at high energies.
In addition to the Regge trajectories, Regge cuts with natural and un-natural parities were included according 
to the ideas developed in Ref.~\cite{DoKa} for pion photoproduction. 
Nucleon resonances in the $s$ channel were parameterized with Breit-Wigner shapes. 
The new model was fitted to data from both $\eta$ and $\eta'$ photoproduction on protons.
In addition to the new cross sections presented in this letter, data from 
\,\cite{GRAAL_2007, CLAS_2009, CBELSA_2009, A2MAMI_2014, GRAAL_2015, CLAS_2016} were used.
A detailed publication of the model including a quantitative comparison to all available data is in preparation.
Here we concentrate on the comparison
to the new cross section data.
A key role for the description  is played by the three $s$-wave resonances, N(1535)$1/2^-$,
N(1650)$1/2^-$, and N(1895)$1/2^-$. The importance of the first two resonances in $\eta$ 
photoproduction is well known from previous analyses.
In our model, the third resonance, N(1895)$1/2^-$, is crucial in order to describe the 
cusp observed in $\eta$ photoproduction around $W=1896$~MeV as well as the fast rise of the total cross section of the $\gamma p \to \eta' p$ reaction near the threshold. 
Presently, this resonance has only an overall two-star status according to the PDG 
review~\cite{PDG}. The present data and our analysis clearly confirm the existence 
of this state.
We find a mass slightly below the $\eta'$ threshold and a significant coupling to both,
$\eta p$ and $\eta' p$. The parameters of all s-wave resonances are presented in Table I.
As the  N(1895)$1/2^-$ mass is below $\eta' N$ threshold, an effective 
branching ratio of $\beta_{\eta' N}=(38 \pm 20)\%$ was determined by integrating 
the decay spectrum above $\eta' N$ threshold according to \cite{pdg2012}. The contributions of this and the other two
important resonances, the $N(1990)1/2^+$  and the $N(2020)3/2^-$, are shown in Fig.~\ref{fig5etapr}.
\begin{table}
\caption{Fit results for $J^P = 1/2^-$ resonances. 
Breit-Wigner parameters: mass M$_{BW}$, width $\Gamma_{BW}$, branching ratio to 
$\eta$N channel $\beta_{\eta N} := \Gamma_{\eta N}(M_{BW})/\Gamma_{BW}$, and helicity amplitude $A_{1/2}$ in $[10^{-3}$GeV$^{-1/2}]$. 
Stars in the first column indicate an overall status of the resonance.
The first row for each resonance gives a parameter set of the new $\eta$MAID solution .
The second row lists the corresponding numbers given by the PDG review~\cite{PDG}. 
The parameters indicated without errors were fixed during the fit.
}
\begin{center} 
\begin{tabular*}{8.6cm}
{@{\hspace{0.01cm}}c @{\hspace{0.05cm}}|
@{\hspace{0.05cm}}c
@{\hspace{0.27cm}}c
@{\hspace{0.27cm}}c
@{\hspace{0.27cm}}c }
\hline\hline\noalign{\smallskip}
Resonance$J^P$ &
M$_{BW}$[MeV] &
$\Gamma_{BW}$[MeV] &
$\beta_{\eta N}[\%]$ &
$A_{1/2}$ \\
\noalign{\smallskip}\hline\noalign{\smallskip}
N(1535)$1/2^-$ & $1528\pm6$  & $163\pm25$  & $41\pm4$ & $+115$ \\
 ****           & $1535\pm10$ & $150\pm25$ & $42\pm10$ & $+115\pm15$ \\
\noalign{\smallskip}\hline\noalign{\smallskip}
N(1650)$1/2^-$ & $1634\pm5$  & $128\pm16$ & $28\pm11$   & $+45$    \\
 ****    & $1655^{+15}_{-10}$ & $140\pm30$ & $14-22$    & $+45\pm10$  \\
\noalign{\smallskip}\hline\noalign{\smallskip}

N(1895)$1/2^-$ &$1890^{+9}_{-23}$ &$150\pm57$ & $20\pm6$   & $-30$  \\
 **                                                                                 \\
\noalign{\smallskip}\hline\hline
\end{tabular*}  
\end{center}   
\end{table}

 In summary, photoproduction reactions $\gamma p\to \eta p$ and $\gamma p\to \eta' p$
 have been measured from their thresholds up to the center-of-mass energy $W=1.96$~GeV
 with the A2 tagged-photon facilities at the Mainz Microtron, MAMI.
 Differential cross sections were obtained with unprecedented statistical accuracy,
 providing fine energy binning and full production-angle coverage.
The total cross section and the excitation functions for $\eta$ photoproduction demonstrate 
a strong cusp in the vicinity of the $\eta'$ threshold, $W=1896$~MeV.
 The analysis of the present data with the revised $\eta$MAID model explains
 such a behavior by the strong coupling of the $N(1895)1/2^-$ resonance to both 
 channels. The new data and our analysis clearly confirm 
 the existence of this two-star state and allow significant improvements
in the determination of its parameters.

The authors wish to acknowledge the excellent support of the accelerator group of MAMI.
This material is based upon work supported by
the Deutsche Forschungsgemeinschaft (SFB 1044),
the European Community Research Activity under the FP7 program (Hadron Physics, Contract No. 
227431),
Schweizerischer Nationalfonds,
the UK Sciences and Technology Facilities Council (STFC 57071/1, 50727/1),
the U.S. Department of Energy (Offices of Science and Nuclear Physics,
Award Nos. DE-FG02-99-ER41110, DE-FG02-88ER40415, DE-FG02-01-ER41194 and DE-FG02-SC0016583)
and National Science Foundation (Grant No. PHY-1039130, IIA-1358175),
NSERC FRN:  the MSE Program ``Nauka'' (Project 3.1113.2017/pp).


\end{document}